\documentclass{article}
\usepackage{spconf,amsmath,graphicx}
\usepackage{hyperref}
\usepackage{cite,url,booktabs,subfig,bm,amssymb,multirow} 

\usepackage{xspace}
\newcommand{\fo}{$F_0$\xspace}

\newcommand{\noAtt}{\textbf{NoAtt}\xspace}
\newcommand{\Prop}{\textbf{Prop}\xspace}
\newcommand{\Base}{\textbf{Base}\xspace}
\newcommand{\NF}{\textbf{NF}\xspace}
\newcommand{\NP}{\textbf{NP}\xspace}
\newcommand{\NPNF}{\textbf{NP+NF}\xspace}
\newcommand{\noTrans}{\textbf{NoTrans}\xspace}
\newcommand{\pTrans}{\textbf{P-Trans}\xspace}
\newcommand{\tTrans}{\textbf{T-Trans}\xspace}

\newcommand{\nb}{\nolinebreak}

\title{Singing Voice Synthesis Based on a Musical Note Position-aware\\ Attention Mechanism}
\name{Yukiya Hono, Kei Hashimoto, Yoshihiko Nankaku, and Keiichi Tokuda
\thanks{This work was supported by JSPS KAKENHI Grant Number JP22H03614, CASIO SCIENCE PROMOTION FOUNDATION, and FOUNDATION OF PUBLIC INTEREST OF TATEMATSU.}}
\address{Nagoya Institute of Technology, Nagoya, Japan}
\begin{document}
\ninept
\maketitle
\begin{abstract}
  This paper proposes a novel sequence-to-sequence (seq2seq) model with a musical note position-aware attention mechanism for singing voice synthesis (SVS).
  A seq2seq modeling approach that can simultaneously perform acoustic and temporal modeling is attractive.
  However, due to the difficulty of the temporal modeling of singing voices, many recent SVS systems with an encoder-decoder-based model still rely on explicitly on duration information generated by additional modules.
  Although some studies perform simultaneous modeling using seq2seq models with an attention mechanism, they have insufficient robustness against temporal modeling.
  The proposed attention mechanism is designed to estimate the attention weights by considering the rhythm given by the musical score.
  Furthermore, several techniques are also introduced to improve the modeling performance of the singing voice.
  Experimental results indicated that the proposed model is effective in terms of both naturalness and robustness of timing.
\end{abstract}
\begin{keywords}
  Singing voice synthesis, sequence-to-sequence model, attention mechanism, temporal modeling
\end{keywords}

\vspace{-1mm}
\section{Introduction}
\vspace{-2mm}
\label{sec:intro}

Statistical parametric singing voice synthesis (SVS) have been evolving with the spread of machine learning techniques~\cite{oura-2010-recent,hono-2021-sinsy,blaauw-2017-neural,hono-2019-singing,nakamura-2020-fast,yi-2019-singing}.
The essence of SVS is a sequence transform to generate an acoustic feature sequence from a score feature sequence obtained from musical scores.
Since the singing voice needs to be strictly synchronized with the given musical score, how to model the temporal structure of the singing voice is one of the important issues in the SVS.

Typical deep neural network (DNN)-based SVS systems~\cite{hono-2021-sinsy,blaauw-2017-neural,hono-2019-singing,nakamura-2020-fast} model the acoustic feature and its temporal structure (specifically as phoneme duration and vocal timing deviation) independently using acoustic, duration, and time-lag models.
The acoustic model works as a mapping function that generates the acoustic feature sequence from the time-aligned score feature sequence.
Although these pipeline systems are stable when running in both training and synthesis, they suffer from alignment-related issues: 1) the modeling performance of acoustic features is affected by alignment errors, and 2) they do not have the sufficient ability to adequately model the correlation between the acoustic feature and its temporal structure.
These errors may cause a lack of naturalness and expressiveness of synthesized singing voices.

Inspired by the success of autoregressive (AR) and non-AR generation models for modern text-to-speech (TTS) systems~\cite{sotelo-2017-char2wav,wang-2017-tacotron,shen-2018-natural,zhang-2018-forward,li-2019-neural,ren-2020-fastspeech2,shen-2020-non,elias-2021-parallel}, several neural SVS systems have been developed~\cite{lee-2019-adversarially,blaauw-2020-sequence,lu-2020-xiaoicesing,chen-2020-hifisinger,wu-2020-adversarially,gu-2021-bytesing,shi-2021-sequence,angelini-2020-singing,wang-2022-singing}.
Unlike TTS, it is not easy to model the temporal structure of a singing voice because the phoneme duration of singing voices varies greatly depending on the corresponding note duration, even for the same phoneme.
Thus, these kinds of SVS systems~\cite{lee-2019-adversarially,blaauw-2020-sequence,lu-2020-xiaoicesing,chen-2020-hifisinger,wu-2020-adversarially,gu-2021-bytesing,shi-2021-sequence} mainly adopt encoder-decoder models with an explicit length regulator.
This approach has the advantage of robustness in terms of alignment while it makes the model sensitive to the performance of external duration information, similar to conventional pipeline systems.
A number of studies~\cite{angelini-2020-singing,wang-2022-singing} use an AR sequence-to-sequence (seq2seq) model with an attention mechanism to model the acoustic feature and its temporal structure simultaneously; however, they suffer from timing mismatches between the target musical score and synthesized singing voice.
This is a critical issue for SVS applications because manually collecting the timing in these attention-based systems is both difficult and impractical

In this paper, we propose a novel seq2seq model for SVS with a musical note position-aware attention mechanism.
The proposed attention mechanism calculates the attention weights based on the note position informed by the musical score.
Moreover, we also introduce additional techniques to help obtain robust alignment and improve naturalness: auxiliary note feature embedding, guided attention loss with a penalty matrix designed for singing voice alignments, and pitch normalization for the seq2seq model.
The proposed seq2seq model can synthesize singing voices with proper vocal timing without extra supplementary temporal modeling.

\vspace{-1mm}
\section{Related work}
\vspace{-2mm}

Seq2seq modeling using an attention mechanism~\cite{bahdanau-2015-neural} is a key technique for modeling the correspondence between sequences with different lengths.
In the TTS scenario, since the monotonicity and locality properties of TTS alignment can be exploited, hybrid location-sensitive attention combining content-based and location-based attention mechanisms~\cite{shen-2018-natural}, a forward attention mechanism computed recursively using a recursively forward algorithm~\cite{zhang-2018-forward}, and monotonic attention mechanisms~\cite{he-2019-robust,yasuda-2019-initial} have been developed.

One of the noticeable differences between SVS and other typical seq2seq tasks is that the alignment path is strongly related to the note timing.
The authors~\cite{wang-2022-singing} have been proposed a global duration control attention, introducing a global transition token into a forward attention mechanism.
Although the tempo of singing voices can be controlled through its token, a time-invariant token has the insufficient ability for duration control and cannot prevent the mismatch between the timing of synthesized singing voices and musical scores.
The proposed musical note position-aware attention mechanism is based on the generalized form of forward attention with phoneme-dependent and time-variant transition probabilities, and the attention probability is calculated taking into account note position information via note position embeddings.
This achieves appropriate modeling of the temporal structure of the singing voice by the attention mechanism.

\begin{figure}[t]
  \centering
  \includegraphics[width=0.85\linewidth]{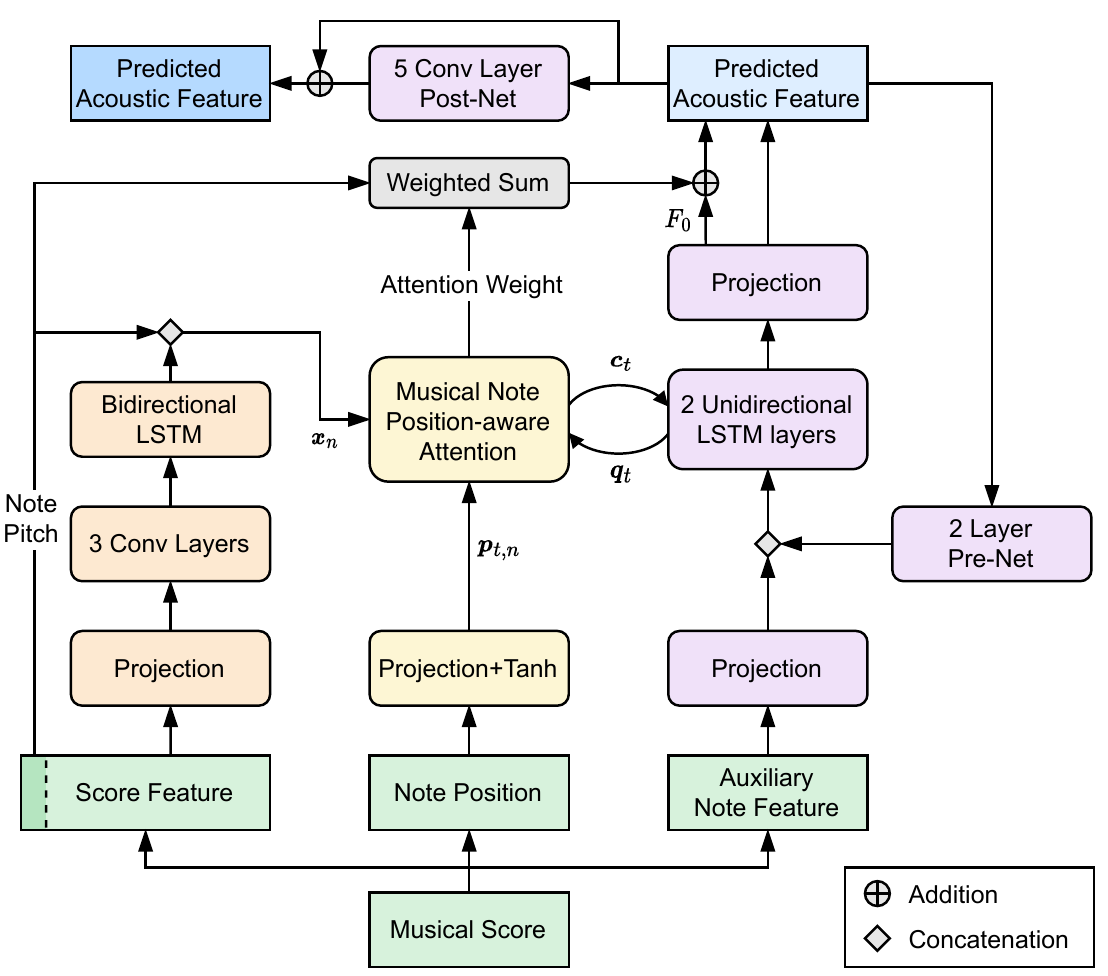}
  \caption{Overview of the proposed model.}
  \vspace{-1mm}
  \label{fig:proposed}
\end{figure}

\vspace{-1mm}
\section{Proposed model}
\label{sec:proposed}
\vspace{-2mm}

The proposed model is based on an encoder-decoder model with an attention mechanism and can generate frame-level acoustic feature sequences directly from phoneme-level score feature sequences.
The score feature consists of not only the phone, note pitch, and note length but also other rich musical contexts such as beat, key, tempo, dynamics, and staccato~\cite{oura-2010-recent}.
An overview of the proposed model is shown in Fig.~\ref{fig:proposed}.
We describe four techniques in this section for seq2seq SVS systems to satisfy the singing-specific requirements, such as high controllability and robustness against tempo and pitch.

\vspace{-1mm}
\subsection{Musical note position-aware attention mechanism}
\label{sec:note-pos-att}
\vspace{-2mm}

The attention mechanism~\cite{bahdanau-2015-neural} calculates an attention weight at each decoder time-step to perform a soft-selection of encoder hidden state $\bm{x} = [\bm{x}_1, \bm{x}_2, \ldots \bm{x}_N]$, which is obtained by processing the input score features by the encoder.
The context vector $\bm{c}_t$ can be obtained by $\bm{c}_t = \sum_{n=1}^N \alpha_t(n) \bm{x}_n$, where $\alpha_{t}(n)$ is the attention weight representing the degree of attention on the $n$-th hidden state at the $t$-th decoder time-step.
Finally, the output vector $\bm{o}_t$ can be the predicted conditioning of the context vector \nb$\bm{c}_t$ by the decoder.

The alignment obtained by the attention mechanism must be monotonic and continuous without skipping any encoder states to synthesize a singing voice following a given musical score.
To satisfy this assumption, a current attention weight $\alpha_t(n)$ is calculated recursively using the previous alignment as follows:
\begin{align}
  \alpha'_t(n) & = \big( (1 - u_{t-1}(n))\alpha_{t-1}(n) \nonumber                                \\
               & \qquad\quad + u_{t - 1}(n - 1)\alpha_{t-1}(n-1)\big) \cdot y_t(n), \label{eq:fa} \\
  \alpha_t(n)  & = \alpha'_t(n) \biggl/\, \sum_{m=1}^N \alpha'_t(m), \label{eq:alpha}
\end{align}
where $y_t(n)$ is the output probability, and $u_{t}(n)$ is the transition probability where the attention mechanism notices the $n$-th phoneme at the $t$-th decoder step and moves to the $n+1$-th phoneme at the $t+1$-th decoder steps.
Note that $u_{t}(n)$ in Eq.~\eqref{eq:fa} is the phoneme-dependent time-variant transition probability; thereby this formula can be regarded as a generalized form of forward attention with a transition agent~\cite{zhang-2018-forward}.

The alignment of SVS should be obtained by considering the temporal structure of musical notes, specifically note duration and tempo.
Therefore, we introduce musical note positional features to compute $y_t(n)$ and $u_{t}(n)$.

In the proposed method, the output probability $y_t(n)$ is calculated by extended content-based attention with a musical note position-aware additional term $\bm{U}^{(\cdot)}\bm{p}_{t,n}$ as:
\begin{align}
  e_t(n) & = \bm{v}^{(e)}\!^\top\! \tanh(\bm{W}^{(e)}\bm{q}_t + \bm{V}^{(e)}\bm{x}_n + \bm{U}^{(e)}\bm{p}_{t,n} + \bm{b}^{(e)}), \\
  y_t(n) & = \exp\left(e_t(n)\right) \biggl/\, \sum_{m=1}^N \exp\left(e_t(m)\right), \label{eq:outprob}
\end{align}
where $\bm{q}_t$ denotes the $t$-th time-step decoder hidden state, and $\bm{W}^{(\cdot)}\bm{q}_t$ and $\bm{V}^{(\cdot)}\bm{x}_n$ are the content-based terms that represent query/key comparisons in the attention mechanism, and $\bm{b}^{(\cdot)}$ is the bias term.
The note position embedded feature $\bm{p}_{t,n}$ is obtained by feeding the note position representation $[p^{1}_{t,n}, p^{2}_{t,n}, p^{3}_{t,n}]$ with a single tanh hidden layer.
Each note position representation is computed from the note lengths of the given musical score as follows:
\begin{align}
  p^{1}_{t, n} & = t - s_n, \\
  p^{2}_{t, n} & = e_n - t, \\
  p^{3}_{t, n} & = \left\{
  \begin{aligned}
     & s_n - t,  & \quad (t < s_n)           \\
     & 0,        & \quad (s_n \le t \le e_n) \\
     & t - e_n,  & \quad (e_n < t)           \\
  \end{aligned} \label{eq:guide}
  \right.
\end{align}
where $s_n$ and $e_n$ denote the start and end positions of the $n$-th musical note, respectively.

Since transition probabilities should be explicitly derived from past alignments, we adopt a location-sensitive attention~\cite{shen-2018-natural}-like formula to calculate the transition probability as follows:
\begin{align}
  u_t(n) = \sigma \bigl( \bm{v}^{(u)}\!^\top\! \tanh(\bm{W}^{(u)}\bm{q}_t &+ \bm{V}^{(u)}\bm{x}_n + \bm{U}^{(u)}\bm{p}_{t,n} \nonumber \\
         &+ \bm{T}^{(u)}\bm{f}_{t, n} + \bm{b}^{(u)}) \bigr), \label{eq:transprob}
\end{align}
where $\sigma(\cdot)$ is a logistic sigmoid function, and $\bm{T}^{(u)}\bm{f}_{t, n}$ denotes the location-sensitive term that uses convolutional features computed from the previous cumulative alignments~\cite{chorowski-2015-attention}.

\vspace{-1mm}
\subsection{Auxiliary note feature embedding}
\label{sec:note-aux-feat}
\vspace{-2mm}

As the temporal structure of the singing voices depends on the context of the notes in the input score, the alignment of singing voices should also be predicted on the basis of it.
To encourage this, we embed musical note context associated with the current note as an auxiliary note feature into the attention query.

The auxiliary note features contained musical note-related contexts, which were obtained by removing phoneme and mora-related contexts from the score features, and were expanded to the frame-level sequence using note length.
This upsampled feature is then fed to a single dense layer, which is concatenated with the output of Pre-Net to form the decoder input.
Since the auxiliary note feature delivers the attention mechanism to the current frame position and note context in the corresponding musical note via the attention query, it is expected to enable the attention mechanism to adjust the alignment to fit the rhythm provided by the musical score.

\begin{figure}[t]
  \centering
  \includegraphics[width=0.8\linewidth]{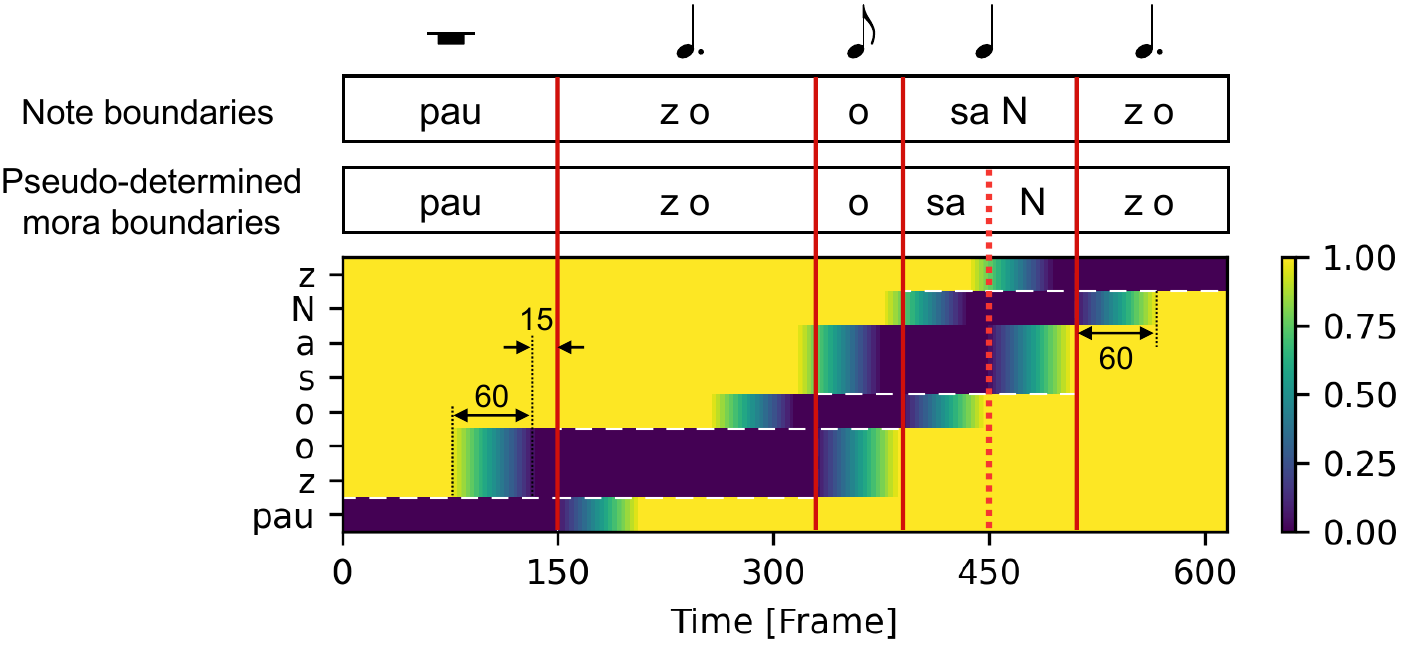}
  \vspace{-2mm}
  \caption{Penalty matrix examples. This matrix is generated on the basis of pseudo-determined mora boundaries from note durations described in the musical score.}
  \vspace{-3mm}
  \label{fig:guide}
\end{figure}

\vspace{-1mm}
\subsection{Guided attention loss for SVS}
\label{sec:mora-guide}
\vspace{-2mm}

Since singing voices are generally sung to follow the rhythm of a musical score, it is natural to assume that the alignment of the singing voice should be close to the path determined by the note timing in the score.
On the basis of this idea, we customize the guided attention loss~\nb\cite{tachibana-2018-efficiently} for SVS systems.

The difference from the original guided attention loss for TTS is how to construct the penalty matrix.
We generate a penalty matrix based on the note boundaries, unlike the one for the TTS whose diagonal elements are zero.
Specifically, we design a penalty matrix $\bm{G} \in \mathbb{R}^{N \times T}$ based on the pseudo-determined mora boundary so that alignment estimation is robust even when multiple morae are included in the same note, as shown in Fig.~\ref{fig:guide}.
These pseudo-boundaries are obtained by equally dividing note duration in accordance with the number of morae in each note.
The soft matrix is generated by linearly decaying over a width of 60 frames.
Note that the start positions of boundaries for constructing the penalty matrix are shifted 15 frames earlier to consider vocal timing deviation.

Let the alignment matrix $\bm{A} \in \mathbb{R}^{N \times T}$ be a matrix, where the element at $(i, j)$ corresponds to $\alpha_j(i)$ and $T$ is the total number of frames.
The guided attention loss is defined as follows:
\begin{align}
  \mathcal{L}_{\mathrm{att}} (\bm{G}, \bm{A}) & = \frac{1}{NT} \left\lVert \bm{G} \odot \bm{A} \right\rVert _1,
\end{align}
where $\odot$ denotes an element-wise product. 
The final loss function $\mathcal{L}$ of the proposed model is given by
\begin{align}
  & \mathcal{L}  = \mathcal{L}_{\mathrm{feat}}(\bm{o}, \hat{\bm{o}}) + \mathcal{L}_{\mathrm{feat}}(\bm{o}, \hat{\bm{o}}') + \lambda \mathcal{L}_{\mathrm{att}} (\bm{G}, \bm{A}), \label{eq:loss} \\
  & \mathcal{L}_{\mathrm{feat}}(\bm{o}, \hat{\bm{o}}) = \frac{1}{TD} \sum_{t=1}^T \left\lVert \bm{o} - \hat{\bm{o}} \right\rVert ^2_2,
\end{align}
where $\hat{\bm{o}}$ and $\hat{\bm{o}}'$ represent the acoustic features predicted by the decoder and Post-Net, respectively, and $D$ is the number of dimensions in the acoustic features.
In Eq.~\eqref{eq:loss}, $\lambda$ represents an adjustment parameter for guided attention loss.

\vspace{-1mm}
\subsection{Pitch normalization}
\label{sec:pitchnorm}
\vspace{-2mm}

The pitch of the synthesized singing voice must accurately follow the note pitch of the musical score.
Thus, following our previous work~\cite{hono-2021-sinsy}, we integrate pitch normalization, where the log fundamental frequency (\fo) is modeled as the difference from the log \fo determined by the musical score (note pitch), into the proposed model.

Pitch normalization requires a time-aligned frame-level note pitch sequence to process the generated \fo sequence frame-by-frame.
In the proposed system, the frame-level note pitch sequence can be obtained by weighting the phone-level input note pitch sequence using the attention weight at each decoder time-step.

\vspace{-1mm}
\section{Experiments}
\label{sec:exp}
\vspace{-2mm}

\subsection{Experimental conditions}
\vspace{-2mm}

To evaluate the proposed models, we conducted experiments using 70 Japanese children's songs (total: 70 min) performed by one female singer.
Sixty songs were used for training, and the rest were used for testing.
Singing voice signals were sampled at 48 kHz, and each sample was quantized by 16 bits.
The acoustic feature consisted of 0-th through 49-th mel-cepstral coefficients, a continuous log \fo value, 25-dimensional analysis aperiodicity measures, 1-dimensional vibrato component, and a voiced/unvoiced binary flag.
Mel-cepstral coefficients were extracted by WORLD~\cite{morise-2016-world}.
The difference between the original log \fo and the median-smoothed log \fo were used as the vibrato component.

The model architectures of encoder, decoder, Pre-Net, and Post-Net are the same as those of \cite{shen-2018-natural}.
A linear projection layer was used instead of the embedding layer.
We added an extra linear projection layer to process the frame-level auxiliary note feature sequence.
The score feature was a 267-dimensional feature vector.
The auxiliary note feature was an 87-dimensional feature vector. 
The frame position in the current note and the note duration were concatenated with an expanded frame-level auxiliary note feature.
For all systems, the reduction factor was set to $3$, and pitch normalization was applied for \fo modeling.
The hyperparameter $\lambda$ in Eq.~\eqref{eq:loss} was set to $10.0$.
All systems were combined with the same pre-trained PeriodNet~\cite{hono-2021-periodnet}, a non-AR neural vocoder with a parallel structure, to reconstruct waveforms from predicted acoustic features.

\vspace{-1mm}
\subsection{Subjective evaluation}
\vspace{-2mm}

\renewcommand{\arraystretch}{0.92}
\begin{table}[t]
  \caption{Results of experiment 1 with 95\% confidence intervals.}
  \vspace{-3mm}
  \label{tb:exp1}
  \centering
  \small
  \begin{tabular}{c|ccc|c}
    \toprule
    \multirow{2}{*}{Systems} & $\bm{p}_{t,n}$                  & \hspace{-3mm}Aux.\! note feat.\hspace{-3mm} & $\mathcal{L}_{\mathrm{att}}$  & \multirow{2}{*}{MOS} \\
                             & (Sec.\! \ref{sec:note-pos-att}) & (Sec.\! \ref{sec:note-aux-feat})            & (Sec.\! \ref{sec:mora-guide}) &                      \\
    \midrule
    \midrule
    \Base              &              &              &              & failed                   \\
    \NF                &              & $\checkmark$ &              & failed                   \\
    \NP                & $\checkmark$ &              &              & $3.12 \pm 0.13$          \\
    \NPNF\hspace{-1mm} & $\checkmark$ & $\checkmark$ &              & $3.81 \pm 0.12$          \\
    \midrule
    \Prop              & $\checkmark$ & $\checkmark$ & $\checkmark$ & \textbf{3.95 $\pm$ 0.12} \\
    \bottomrule
  \end{tabular}
  \vspace{-2mm}
\end{table}
\renewcommand{\arraystretch}{1.0}

We performed 5-scale mean opinion score (MOS) tests\footnote{Audio samples are available at the following URL: \url{https://www.sp.nitech.ac.jp/~hono/demos/icassp2023/}} to evaluate the naturalness of the synthesized singing voices.
Each of the 15 native Japanese-speaking participants evaluated ten phrases randomly selected from the test songs.
A click sound generated based on the basis of the tempo of the score was superimposed on the synthesized singing voice to evaluate the overall naturalness considering the vocal timing.

\vspace{-1mm}
\subsubsection{Experiment 1}
\vspace{-2mm}

\begin{figure}[t]
  \centering
  \vspace{-4mm}
  \subfloat[\Base]{\includegraphics[width=0.41\hsize]{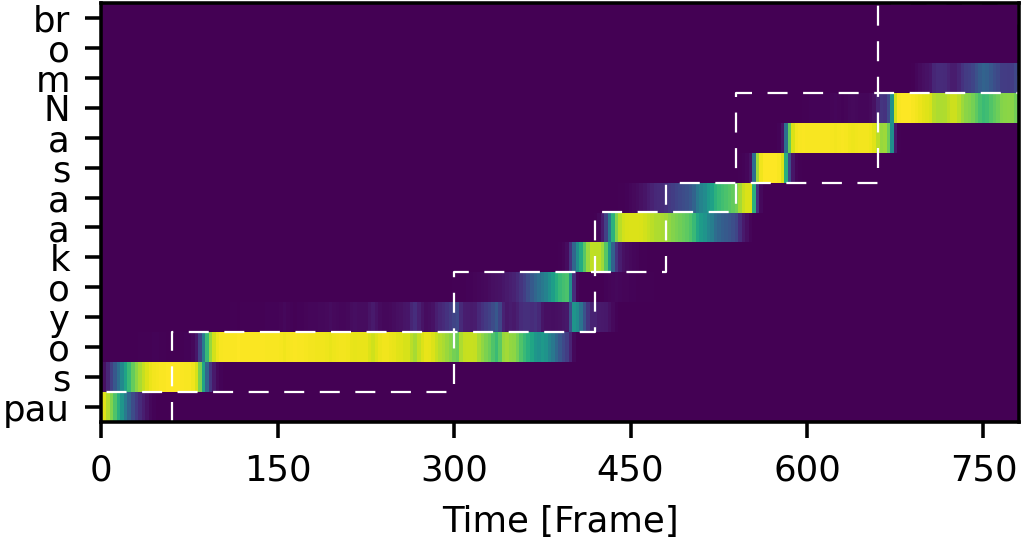}}
  \hspace{4mm}
  \subfloat[\NF]{\includegraphics[width=0.41\hsize]{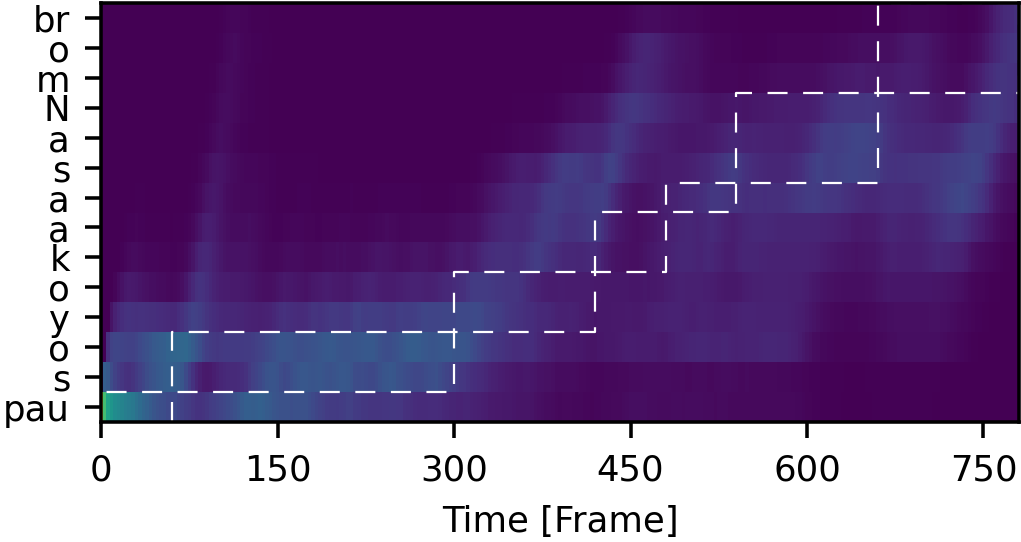}}
  \\
  \vspace{-2mm}
  \subfloat[\NP]{\includegraphics[width=0.41\hsize]{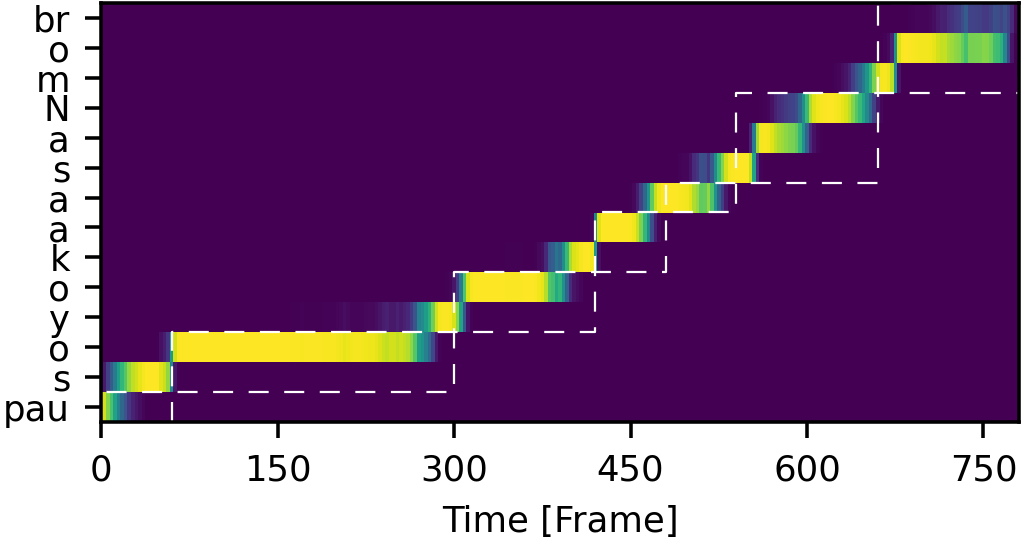}}
  \hspace{4mm}
  \subfloat[\NPNF]{\includegraphics[width=0.41\hsize]{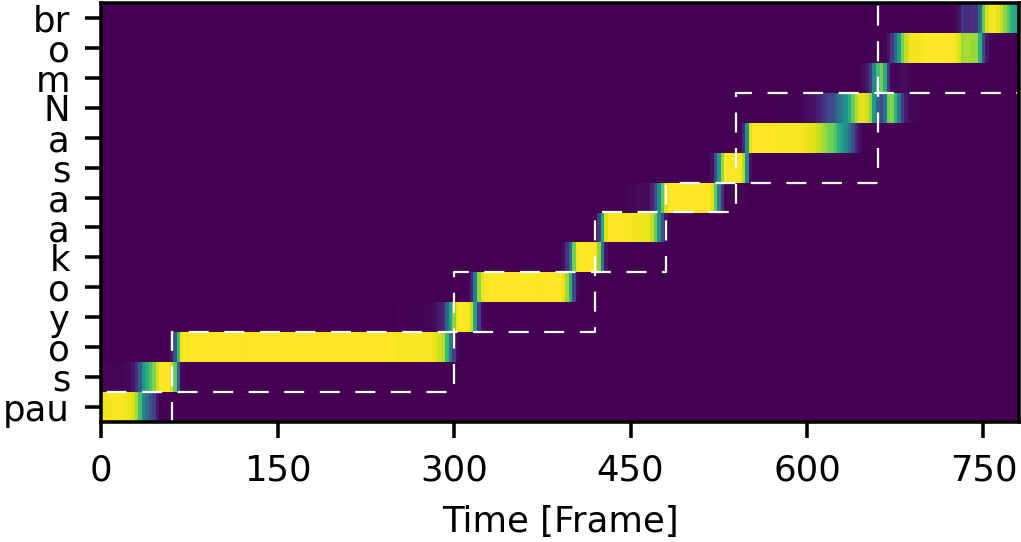}}
  \\
  \vspace{-2mm}
  \subfloat[\noTrans \label{fig:noTrans-att}]{\includegraphics[width=0.41\hsize]{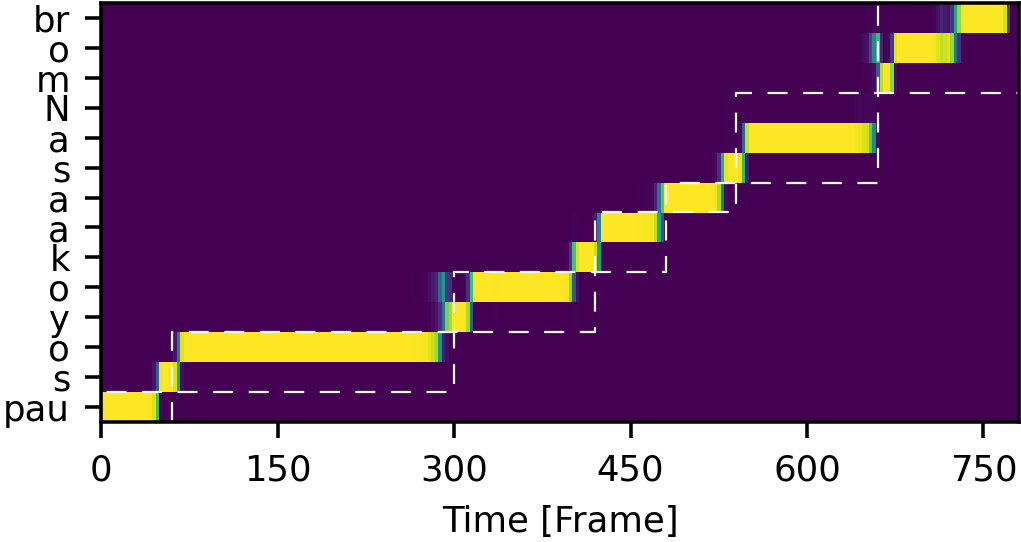}}
  \hspace{4mm}
  \subfloat[\pTrans \label{fig:pTrans-att}]{\includegraphics[width=0.41\hsize]{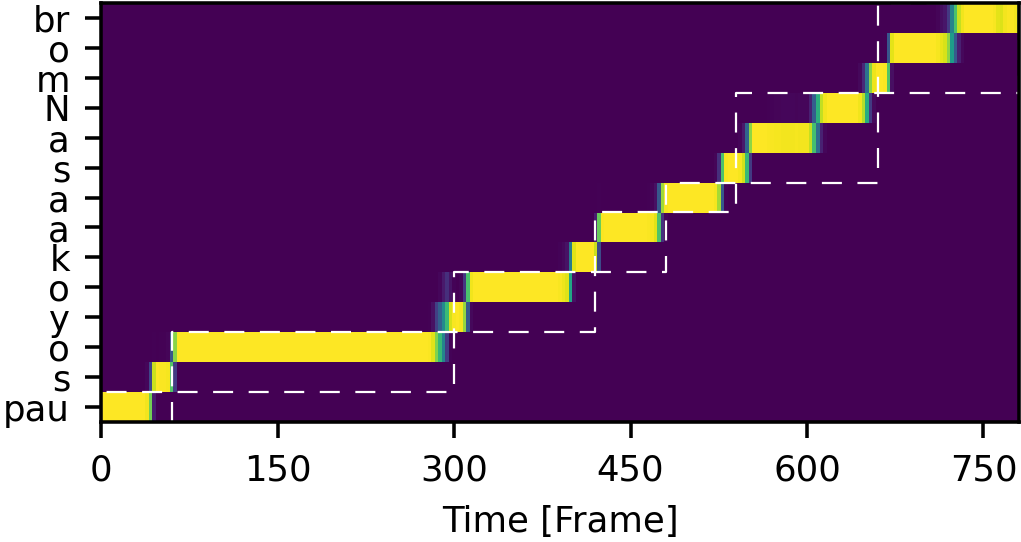}}
  \\
  \vspace{-2mm}
  \subfloat[\tTrans \label{fig:tTrans-att}]{\includegraphics[width=0.41\hsize]{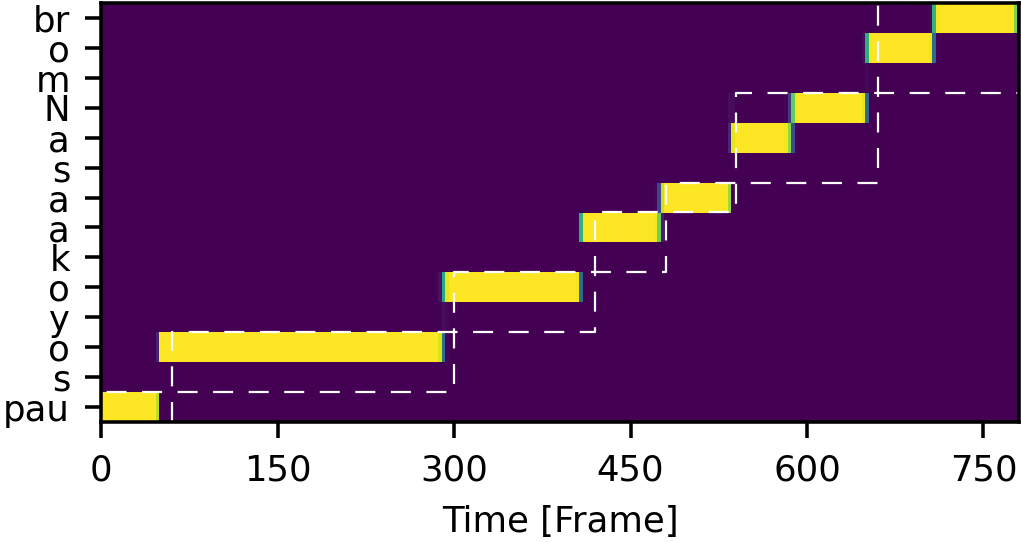}}
  \hspace{4mm}
  \subfloat[\Prop \label{fig:prop-att}]{\includegraphics[width=0.41\hsize]{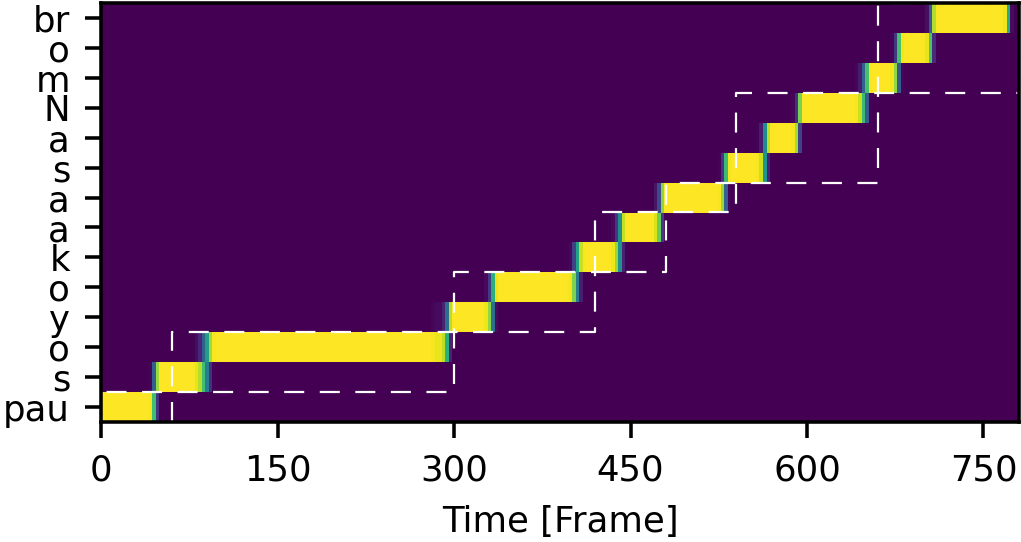}}
  \vspace{-2mm}
  \caption{Examples of alignments obtained by each system.
    White dashed lines represent note boundaries.}
  \vspace{-1mm}
  \label{fig:att}
\end{figure}

\begin{figure}[t]
  \centering
  \vspace{-2mm}
  \subfloat[\pTrans \label{fig:pTrans-ot}]{\includegraphics[width=0.82\hsize]{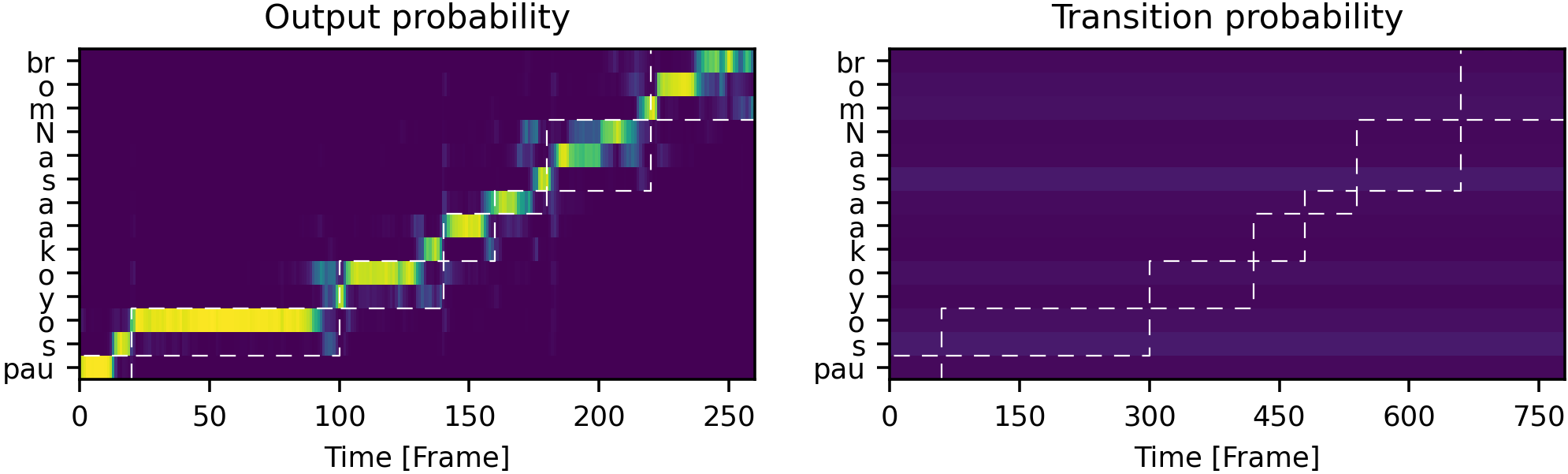}}
  \vspace{-2mm}
  \\
  \subfloat[\tTrans \label{fig:tTrans-ot}]{\includegraphics[width=0.82\hsize]{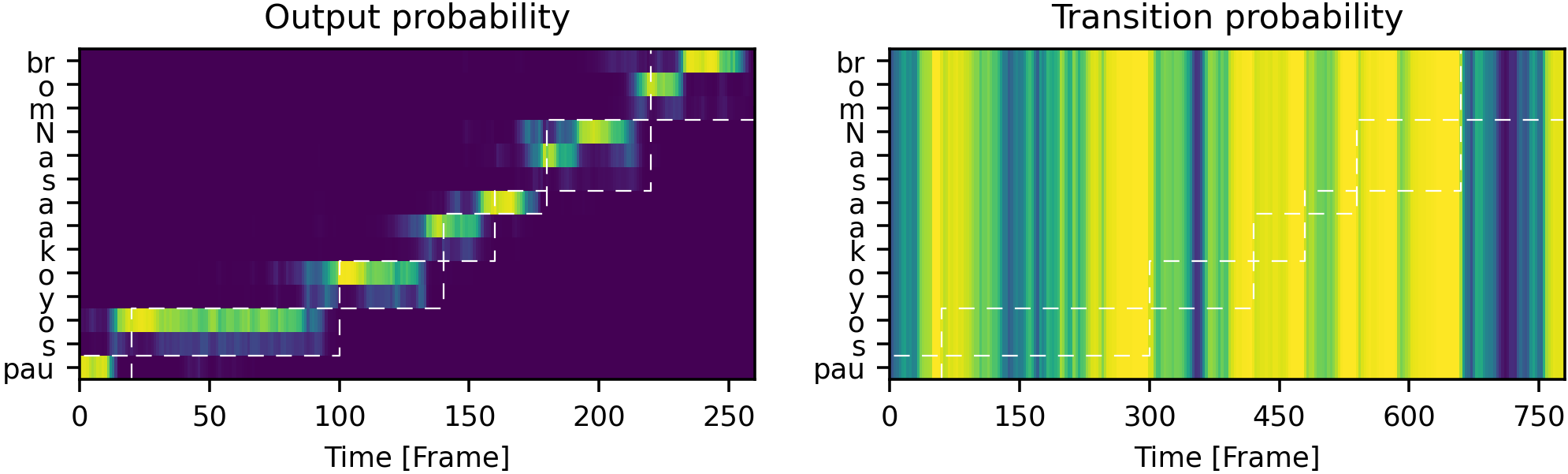}}
  \vspace{-2mm}
  \\
  \subfloat[\Prop \label{fig:prop-ot}]{\includegraphics[width=0.82\hsize]{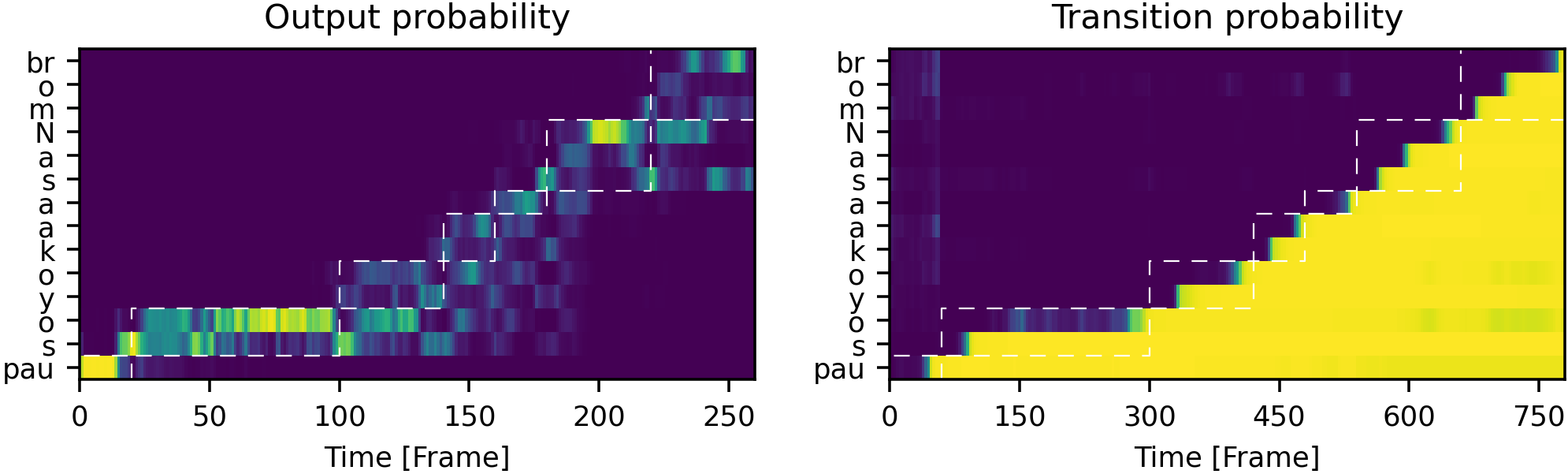}}
  \vspace{-2mm}

  \caption{Examples of output probabilities $y_t(n)$ (left column) and transition probabilities $u_{t}(n)$ (right column) in \pTrans, \tTrans, and \Prop. White dashed lines represent note boundaries.}
  \label{fig:prop}
  \vspace{-3mm}
\end{figure}

We first evaluated the effectiveness of the proposed techniques for modeling the temporal structure of the singing voice described in Section~\nb\ref{sec:proposed}.
We used five systems as shown in Table~\ref{tb:exp1}.
In this experiment, the attention weights were calculated following Eqs.~\nb\eqref{eq:fa} and \eqref{eq:alpha}.
In \Base and \NF, we calculated the output and transition probabilities without using $\bm{U}^{(\cdot)}\bm{p}_{t,n}$ in Eqs.~\nb\eqref{eq:outprob} and \eqref{eq:transprob}.

Table~\ref{tb:exp1} shows the subjective evaluation results, and Fig.~\ref{fig:att} shows a number of examples of the alignments predicted by each system.
From Fig.~\ref{fig:att}, \Base and \NF cannot obtain an appropriate alignment of the singing voice, which indicates the proposed musical note position-aware attention mechanism is effective.
Since synthesized samples by \Base and \NF contained many timing and lyrics errors, they were excluded from the MOS test.
\NPNF achieved much higher MOS scores than \NP.
This shows that the use of auxiliary note features is effective.
The figure also shows that \Prop can obtain a more monotonic alignment and better MOS scores than \NPNF and thus demonstrating the effectiveness of the proposed guided attention loss described in Section~\ref{sec:mora-guide}.
Note that a previous study~\cite{angelini-2020-singing} reported that a simple attention model similar to \Base could synthesize singing voices; however, this required 70 hours of singing voice data with data augmentation.
In contrast, our proposed system does not require large training data or data augmentation approaches.

\vspace{-1mm}
\subsubsection{Experiment 2}
\vspace{-2mm}

\renewcommand{\arraystretch}{0.92}
\begin{table}[t]
  \caption{Results of experiment 2 with 95\% confidence intervals.}
  \vspace{-3mm}
  \label{tb:exp2}
  \centering
  \small
  \begin{tabular}{c|ccc|c}
    \toprule
     & \multirow{3}{*}{\shortstack{\\Attention\\mechanism}}\hspace{-1mm} & \multicolumn{2}{c|}{Transition probability} & \\
    Systems & & \hspace{-1mm}\multirow{2}{*}{\shortstack{\\Phoneme-\\dependent}} & \multirow{2}{*}{\shortstack{\\Time-\\variant}} & MOS \\
                         &              &              &              &                          \\
    \midrule
    \midrule
    \noAtt               &              &              &              & $3.61 \pm 0.13$          \\
    \midrule
    \noTrans             & $\checkmark$ &              &              & $3.78 \pm 0.12$          \\
    \pTrans\hspace{-1mm} & $\checkmark$ & $\checkmark$ &              & $3.83 \pm 0.13$          \\
    \tTrans\hspace{-1mm} & $\checkmark$ &              & $\checkmark$ & $2.16 \pm 0.12$          \\
    \midrule
    \Prop                & $\checkmark$ & $\checkmark$ & $\checkmark$ & \textbf{3.87 $\pm$ 0.12} \\
    \bottomrule
  \end{tabular}
  \vspace{-2mm}
\end{table}
\renewcommand{\arraystretch}{1.0}

We performed another experiment focusing on the attention mechanisms using the five systems listed in Table~\ref{tb:exp2}.
Note that \Prop in Tables \nb\ref{tb:exp1} and \ref{tb:exp2} denote the same systems.
In \noAtt, instead of an attention mechanism, the encoder hidden state is time-aligned by using forced alignment at the training stage and estimating phoneme boundaries using DNN duration and time-lag models at the synthesis stage.
Five-state, left-to-right, no-skip hidden semi-Markov models (HSMMs) were used to obtain forced alignment~\cite{zen-2007-hidden}.
\noTrans is the system where the transition probability is fixed at $u_{t}(n)=0.5$, \pTrans is the system where the transition probability is only phoneme-dependent, and \tTrans is the system where transition probability is time-variant regardless of the phonemes.
The other conditions of \noTrans, \pTrans and \tTrans are the same as those in \Prop.
The attention mechanism of \pTrans is similar to \cite{wang-2022-singing} in exploiting the phoneme-dependent transition probability.

Table~\ref{tb:exp2} and Fig.~\ref{fig:att} show the subjective evaluation results and examples of the alignments predicted by each system, respectively.
\pTrans had a slightly better score than \noTrans.
As shown in Fig.~\ref{fig:pTrans-att}, the alignment obtained by \pTrans increased monotonically; however, it can be seen from Fig.~\ref{fig:pTrans-ot} that estimating the alignments is highly dependent on the output probabilities.
On the other hand, as shown in Figs.~\ref{fig:tTrans-att} and \ref{fig:tTrans-ot}, the alignment of \tTrans readily undergoes transitions, which causes consonant skipping.
As the attention mechanism of \tTrans cannot model temporal structures appropriately, \tTrans had the lowest MOS scores in the subjective evaluation.
In contrast, \Prop achieved the best MOS score.
Figures~\ref{fig:prop-att} and \ref{fig:prop} show that the transition probability of \Prop works more adequately than others, which leads to obtaining the appropriate alignment.
Furthermore, \Prop achieved a better MOS score than \noAtt.
This indicates the effectiveness of the proposed method whereby acoustic features and temporal structures of the singing voice are modeled simultaneously using the seq2seq model with the proposed attention mechanism.

\vspace{-1mm}
\section{Conclusion}
\vspace{-2mm}

We proposed singing voice synthesis based on the seq2seq model with a musical note position-aware attention mechanism.
Our proposed model can properly estimate the alignment of a singing voice while taking note positions into account and can generate acoustic feature sequence without additional temporal modeling.
Experimental results show that the proposed model is effective for simultaneously modeling the acoustic feature sequence and its temporal structure.
Future work includes introducing an HSMM-based structured attention mechanism~\cite{nankaku-2021-neural} to improve the duration controllability and more robust alignment estimation.

\vspace{-1mm}
\section{Acknowledgments}
\vspace{-2mm}

The authors would like to thank Mr. Shumma Murata for joining the discussion and the experiment.

\vfill\pagebreak

\bibliographystyle{IEEEtran}
\bibliography{references}

\end{document}